\documentclass{ws-mpla}
\usepackage{graphicx}
\usepackage{amssymb}
\begin{document}
\markboth{Antonio Feoli} {The amplitude of the de Broglie
Gravitational Waves}
%%% ----------------------------------------------------------------------

\title{The amplitude of the de Broglie Gravitational Waves}

\author{ANTONIO FEOLI}

\address{
Dipartimento d'Ingegneria, Universit\`{a} del Sannio, \\ Corso
Garibaldi n. 107, Palazzo Bosco Lucarelli - 82100 - Benevento -
Italy \\  feoli@unisannio.it}

\maketitle

\begin{abstract}
We calculate the amplitude of the de Broglie gravitational waves
using the standard Einstein General Relativity. We find that these
waves disappear in the limit $\hbar \rightarrow 0$ and when their
source has a large mass and volume. From the experimental point of
view, the knowledge of the amplitude allows to estimate the
magnitude of the effect of the wave on a sphere of test particles.
We propose also to measure a very special shift angle that does
not change with time.

\end{abstract}

\ccode{ PACS numbers: 04.30.-w,  03.75.-b }

\section{Introduction}
In the standard interpretation of Quantum Mechanics the wave
function is the solution of the Schr\"{o}dinger equation and plays
the role of a mathematical tool in calculating the probability for
finding the particle in a given volume of space. In this framework
the normalization constant of the wave function is determined by
the condition
$$
\int_{V}|\psi|^{2} \mathrm{d}V=1. \eqno(1)
$$
On the contrary, de Broglie interpreted this wave $\psi$ as a
genuine field guiding the particle in spacetime.  In 1924 he
proposed\cite{1,2} the relation $P^\nu = \hbar K^\nu$ between the
classical conserved momentum $P^\nu = mcu^\nu = (E/c, \vec{p})$ of
a free particle and the wave number $K^\nu \equiv (\omega/c,
\vec{k})$ of the associated plane wave. Of course the
normalization of four velocity $u^\nu u_\nu = 1$ implies $P^\nu
P_\nu = m^2c^2$ where $m$ is the rest mass of the particle and
$$ u_{\mu}=\left(\frac{1}{\sqrt{1-\left(v^2/c^2\right)}},
-\frac{v}{c\sqrt{1-\left(v^2/c^2\right)}},0,0\right) \eqno(2)
$$
for a particle moving along the x - axis with a constant velocity
$v$.

Generalizing this concept, in 1927, de Broglie  formulated "the
theory of the double solution"\cite{3} in which the particles are
accompanied in spacetime by a {\it real} "pilot wave" $$ \psi =
R(t,x,y,z)exp \left[ \frac{i}{\hbar} S(t,x,y,z)\right] =
R(t,x,y,z)exp \left[ \frac{i}{\hbar} P_\mu x^\mu\right] \eqno(3)
$$ in such a way that the guiding formula: $\partial^\mu S = -P^\mu$ holds.
 Then this idea was
improved in 1952 by Bohm's hidden variable theory\cite{4,5,6,7}
and by Vigier and his co-workers in the Stochastic Interpretation
of Quantum Mechanics\cite{8,9,10}  that is now an alternative to
the standard point of view. Following this line of thinking, if we
want to believe in the reality of de Broglie waves,
 we must give an answer to Bell's famous question: "What is it that $<<waves>>$ in wave mechanics?"
\cite{11} specifying which genuine field is responsible for the de
Broglie wave and simultaneously determining the amplitude $R$ of
these waves somehow. We stress that in the whole  paper we use the
word "amplitude" in the framework of classical physics as "the
half of a total oscillation between a maximum and a minimum of the
wave" and not with the meaning of "probability amplitude" related
in quantum mechanics to inner products between state vectors. The
hope is that one can find an amplitude that will vanish in the
limit $\hbar\rightarrow0$ and will be very small for macroscopic
bodies with a large mass and extension. A recent proposal consists
in considering the de Broglie waves of one-particle quantum
mechanics as a special kind of Gravitational Waves. A solution of
the linearized Einstein equations found in 1998 by Feoli and
Scarpetta\cite{12} has in fact the right features to be
interpreted as the wave associated to a quantum particle because
the guiding formula is satisfied. A similar solution and
interpretation was also found in 2005 by Chang\cite{13} for a
special kind of Electromagnetic Waves.

Starting from a metric tensor $ g_{\mu\nu} = \eta_{\mu\nu} +
h_{\mu\nu}$ (where $\eta_{\mu \nu} = diag(1,-1,-1,-1)$) and
neglecting any terms of order $ |h_{\mu\nu}|^2$, the linearized
Einstein field equations in vacuo are
 $$
 \partial_\alpha\partial^\alpha h_{\mu\nu} = 0 \eqno(4)
 $$
if the de Donder-Lanczos conditions
$$
\partial_\mu \left( h^{\mu}_{\nu} - \frac{1}{2}\delta^{\mu}_{\nu}h \right) = 0\eqno(5)
$$
 are satisfied. The classical solution is a plane wave with $K_\nu K^\nu = 0$.
On the contrary, we found\cite{12} a solution in the form of a
wave packet of cylindrical symmetry around the propagation
direction:
$$ h_{\mu\nu}= e_{\mu\nu} (K_o, K_1) A J_o(\sqrt{y^2 +
z^2}/\lambda) cos \left( K^0 ct - K^1 x \right) \eqno(6) $$ where
$A$ is an arbitrary constant that in this paper we want to
calculate, $J_0$ is the Bessel function of the 0th-order,
$e_{\mu\nu}$ is the polarization tensor and $K^\mu$, the constant
wave number, is such that $K^\mu K_\mu = \lambda^{-2} $,
corresponding to a gravitational wave propagating along the $x$ -
axis with a phase velocity $ V_{ph} = c\sqrt{1 + (\lambda
k)^{-2}}$. If we put $\lambda = \hbar/mc$, this solution can be
interpreted as the de Broglie wave associated with a particle of
rest mass $m$ moving with constant velocity $v$ along the $x$ -
axis because the phase has the right behavior $K^0 ct - K^1 x =
(P_\nu x^\nu)/\hbar = (Et - px)/\hbar$. As in the standard Quantum
Mechanics, we can add waves with slightly different velocities
around a central value $v_0$ to obtain\cite{14,15} a de Broglie
wave packet with a group velocity $V_G = v_0$ such that $V_{ph}
\cdot V_G = c^2$.

From (5) we obtained two relations among the nonvanishing
components of the polarization tensor that turn out to be
depending on the wave number:
$$ e_{00} = e_{11} = \frac{e_{10}}{2} \left( \frac{K_1}{K_0} +
\frac{K_0}{K_1} \right) \eqno(7) $$ $$ e_{22} = e_{33} =
\frac{e_{10}}{2} \left( \frac{K_1}{K_0} - \frac{K_0}{K_1} \right).
\eqno(8) $$

An interesting limit case is when $K_1 = 0$, i.e. when we have an
oscillation in time but not a propagation of the wave and we are
in the rest frame of the associated particle. In that case we
found $ e_{10} = 0$ (hence $e_{10}$ is proportional to $K_1$, that
is to the velocity of the associated particle $v$) and $ e_{22} =
e_{33} = -e_{00} = -e_{11} $. As in this limit, the only
difference among the diagonal components of the constant
polarization tensor is the sign,  we can write $|e_{\mu\mu}|= q$
where $q$ is a constant to be determined. But, without changing
the solution (6), we can also put $|e_{\mu\mu}|= 1$ and redefine
$A$ as $A^{\prime} = q A$. In this way we have at least fixed the
magnitude of the diagonal components of $e_{\mu\nu}$ in the rest
frame of the associated particle.

 The properties of the de Broglie Gravitational waves were studied in
a series of interesting papers\cite{12,14,15,16}, but the
normalization constant $A$ of these waves has always been left
undetermined. In this letter we propose a way to calculate this
constant using the standard methods of Einstein General Relativity
(section 2). Once the amplitude is completely known,  a test to
verify the existence of this kind of Gravitational Waves becomes
possible. When the wave meets a sphere of test particles, one can
hope to measure not only the magnitude of the shift but also its
direction. To this aim we will calculate (section 3) a very
special shift angle that remains constant in time.
\section{Determination of the amplitude}
Following Misner, Thorne and Wheeler\cite{17}, given the
linearized field equations (4) and assuming the gauge conditions
(5),  we can calculate what they call "the effective smeared--out
stress--energy of gravitational waves", starting from an energy -
momentum tensor in the form:
$$
T_{\mu\nu}^{(GW)}=\frac{c^4}{32 \pi
G}\langle\partial_{\mu}h^{\alpha\beta}\partial_{\nu}h_{\alpha\beta}
-\frac{1}{2}\partial_{\mu} h \, \partial_{\nu}h\rangle \eqno(9)
$$
where $<....>$ denotes an average over several periods in time and
wavelengths in space. This tensor\cite{17}``contributes to the
large--scale background curvature (which linearized theory
ignores) just as any other stress--energy does:
$$
G_{\mu\nu}=\frac{8 \pi
G}{c^4}\left(T_{\mu\nu}^{(GW)}+T_{\mu\nu}^{(\mathrm{matter})}
+T_{\mu\nu}^{(\mathrm{other fields})}\right)". \eqno(10)
$$
In our simple case there are no "other fields" and no other
"matter" except the particle source of our gravitational field (6)
already included in the $T_{\mu\nu}^{(GW)}$. If this particle,
associated with the de Broglie wave, is placed in a small volume
$V$ around $r = 0$, i.e. $ V<<\lambda^3$, the tensor
$T_{\mu\nu}^{(GW)}$ must coincide in the limit $r \rightarrow 0$
with $T_{\mu\nu}^{(\mathrm{matter})}=\rho u_{\mu}u_{\nu}$ so the
Einstein equations including terms of the order $|h_{\mu\nu}|^2$
can be written:
$$G_{\mu\nu}=\frac{8 \pi
G}{c^4} T_{\mu\nu}^{(GW)} \longrightarrow \frac{8 \pi G}{c^4}
T_{\mu\nu}^{(\mathrm{matter})}\,\,\, for \,\,\,r \rightarrow 0.
\eqno(11)$$ This really happens constructing our energy--momentum
tensor from the gravitational field (6). Actually, if we write the
solution (6), that (fixing $\lambda = \hbar/mc$) represents our de
Broglie wave, in the form
$$
h_{\alpha\beta}=e_{\alpha\beta}\phi \eqno(12)
$$
where
$$
\phi=A J_{0} \left(\sqrt{y^2+z^2}/\lambda\right)
\cos\left[\left(u_{0}x^{0}+u_{1}x^{1}\right)/\lambda\right]
\eqno(13)
$$
and $u_{\mu}$ is given by the equation (2), we obtain from (9) the
nonvanishing components of the tensor
$$
T_{\mu\nu}^{(GW)}=\langle \frac{c^{4}(e_{22})^2}{16 \pi G
}\partial_{\mu} \phi \, \partial_{\nu}\phi\rangle \eqno(14)
$$
that are the following:
$$
T_{00}^{(GW)}= \frac{c^4(e_{22})^2 A^2 J_{0}^2}{32 \pi G \lambda^2
}u_{0}u_{0} \eqno(15)
$$
$$
T_{11}^{(GW)}= \frac{c^4(e_{22})^2 A^2 J_{0}^2}{32 \pi G \lambda^2
}u_{1}u_{1} \eqno(16)
$$
$$
T_{01}^{(GW)}= \frac{c^4(e_{22})^2 A^2 J_{0}^2}{32 \pi G \lambda^2
}u_{0}u_{1} \eqno(17)
$$
$$
T_{22}^{(GW)}= \frac{c^4(e_{22})^2 A^2}{32 \pi G \lambda^2
}\frac{y^2 J_{1}^2}{\left(z^2+y^2\right)} \eqno(18)
$$
$$
T_{33}^{(GW)}= \frac{c^4(e_{22})^2 A^2}{32 \pi G \lambda^2
}\frac{z^2 J_{1}^2}{\left(z^2+y^2\right)} \eqno(19)
$$
$$
T_{23}^{(GW)}= \frac{c^4(e_{22})^2 A^2}{32 \pi G \lambda^2
}\frac{yz J_{1}^2}{\left(z^2+y^2\right)} \eqno(20)
$$
In the limit $r\rightarrow0$ we have $J_{0}\rightarrow1$ and
$J_{1}\rightarrow0$ so this tensor reduces to the desired form
$T_{\mu\nu}=\rho u_{\mu}u_{\nu}$ where
$$
\rho=\frac{c^4 A^2 J_{0}^2(0)(e_{22})^2 }{32 \pi G \lambda^2 }
\eqno(21)
$$
that can be interpreted as the energy density of a dust--like
source of our gravitational field.

In the rest frame of the particle $\rho = mc^2/V$ and, as
discussed in the previous section, $|e_{\mu\mu}|= 1$ hence
$(e_{22})^2=1$ and
$$
\frac{m c^2}{V}=\frac{c^4 A^2}{32 \pi G \lambda^2} \eqno(22)
$$
where $V$ is the proper volume of the particle. Finally we  find:
$$
A=\frac{4\lambda}{c}\sqrt{\frac{2 \pi G m}{V}} \eqno(23)
$$
and, fixing $\lambda=\hbar/mc$, we obtain:
$$
A=\frac{4\hbar}{c^2}\sqrt{\frac{2 \pi G}{mV}}. \eqno(24)
$$
From $|h_{\mu\nu}|^2 \ll 1$ we have that our model works if
$$
mV \gg \frac{32 \pi G \hbar^2}{c^4}=9.2\times10^{-111} \,
\mathrm{kg \, m^3}. \eqno(25)$$
For example, for an electron the dust - like approximation of the
energy - momentum tensor  works if $V_e=4\pi R_e^3/3 \ll
\lambda_e^3 = (3.86 \times 10^{-13}m)^3$ and the linear
approximation of Einstein equations does if $V_e \gg 10^{-80}
m^3$. Of course the two relations are satisfied both using the
"classical radius" of the electron $R_e = 2.8 \times 10^{-15}m$,
and with the lower experimental value $R_e \simeq 10^{-22} m$
proposed by Dehmelt\cite{18}. In this last case the constant (24)
would become $A \simeq 4.9 \times 10^{-8}$ (more precisely the
Dehmelt value computed from $R_e = \lambda_e |g - 2|$ is $R_e =
4.3 \times 10^{-23}m$ and the corresponding constant becomes $A =
1.7 \times 10^{-7}$) and it would be more convenient to measure
its effects rather than the ones produced by the standard
gravitational waves.

 Note that, in general, due to the
cylindrical symmetry of the solution (12), the constraint $V = \pi
L (y^2 + z^2) \ll \lambda^3$ can be a little bit relaxed. In order
to obtain the dust - like tensor, it is enough that the transverse
radius of the source $\sqrt{y^2 + z^2} \ll \lambda$, while the
length $L$ in the $x$-direction of the cylindrical tube can be
such that $L > \lambda$. In that case, $\sqrt{y^2 + z^2} \ll L$
and our particle becomes similar to a string.

 Furthermore we stress that the  constant $A$, that in the rest frame fixes the
 magnitude of the amplitude $R = A J_0$, has just the right expected behavior because it vanishes in the
limit $\hbar\rightarrow0$ and is negligible when $m$ and $V$ are
very large. In the standard quantum mechanics a fixed scale
parameter is missing, so the border between classical bodies and
quantum particles is not clear. In our approach, the result (24)
allows at least to say that the linear approximation works when
(25) holds and the quantum effects disappear just when
$\hbar\rightarrow0$ or when the product $mV$ is very large
compared to a fixed parameter: $9.2\times10^{-111} \, \mathrm{kg
\, m^3}$ that specifies  the realm of quantum physics.
\section{Experimental test: a very special "shift angle"}
In order to propose an experimental test we must refer to the
tidal acceleration of geodesic deviation between two particles
that does not depend on the particular gauge chosen. Considering a
central particle $m_0$ in the origin of the reference frame, we
could estimate the shift of any peripheral particle at distance $%
\ell$ from the origin. When the de Broglie gravitational wave
passes, moving along the x-axis, it deforms what was a sphere of
radius $\ell$ of test particles,  as measured in the proper frame
of the central particle. We will calculate explicitly only the
shift of the three particles placed on the three axes at the
position $\ell$ and the opposite will occur for the three
particles at $-\ell$.
  In some previous papers\cite{15,16}
we have already calculated the tidal acceleration between two test
particles using the relation:
$$
\frac{D^2 \eta^{\mu}}{c^2 d\tau^2} = -
R^{\mu}_{\nu\gamma\delta}\beta^{\nu}\beta^{\delta} \eta^{\gamma}
\eqno(26)
$$
where $\tau$ is the proper time along the worldline of the two masses, $%
\beta^{\mu}$ is their four - velocity and $\eta^{\mu}$ their
relative distance, while the velocity of the source is $u_\mu$ in
(2). We can use two approximations: the velocity of the test
particles is very small compared with the speed of light so
$\beta^{\mu} \simeq (1,0,0,0)$ and the relative displacement
$\delta^{\mu}$ between two of these masses caused by the wave is
very small. So, if the initial position of the first particle
$m_0$ is
at the origin and of the second one at $\xi^{\mu} = (0, x, y , z )$, we have $%
\eta^{\mu} = (0, x + \delta^{1}, y + \delta^{2}, z + \delta^{3})$
and we can consider $x^{i} >>\delta^{i}$. In this framework the
acceleration of geodesic deviation is
$$
\ddot{\eta}^{k} \simeq - c^2 R^{k}_{0j0} \xi^j \eqno(27)
$$
where a dot indicates a differentiation with respect to time $t$
and $i,j = 1,2,3$. Starting from (6) we calculate:

 a) The relative distance between the mass $m_0$ at the origin and the mass $m_1$ initially at $\xi^{j} = (\ell, 0, 0)$
$$
\eta_1= \delta_1 + \ell = {\frac{A \ell}{2 K_0^2}} J_0(0)\left[ (K_0^2 + K_1^2)e_{00} - 2K_0 K_1 e_{01}\right] \cos \left({%
K_0 ct +K_1 \ell}\right) + \ell \eqno(28)
$$

b) The relative distance between $m_0$ and $m_2$ initially at
$\xi^{j} = (0, \ell, 0)$
$$
\eta_1 = \delta_1 = {\frac{A \ell}{2 \lambda K_0^2}} \left(K_0
e_{10} - K_1 e_{00}\right) J_1(\ell/\lambda)  \sin \left( K_0
ct\right) \eqno(29)
$$

$$
\eta_2 = \delta_2 + \ell = {\frac{A \ell}{2 K_0^2}} \left\{e_{22} K_0^2 J_0(\ell/%
\lambda) - e_{00} \left({\frac{J_2(\ell/%
\lambda)-J_0(\ell/\lambda)}{2 \lambda^2}}\right)\right\} \cos
\left( K_0 ct\right) + \ell \eqno(30) $$

$$ \eta_3 = 0  \eqno(31)$$

c) The relative distance between $m_0$ and  $m_3$ initially at
$\xi^{j} = (0, 0, \ell)$
$$ (\eta_1)_{m_3} = (\eta_1)_{m_2} \;\;\;  (\eta_2)_{m_3} = 0
 \;\;\; (\eta_3)_{m_3} = (\eta_2)_{m_2} \eqno(32)$$

A remarkable result is the existence of a longitudinal
displacement $\delta_1$. So the sphere of particles surrounding
the central mass $m_0$ is transformed into a triaxial ellipsoid,
as we showed in the figures of a previous paper\cite{15}. This
effect is absent in the standard transverse gravitational
waves\cite{19}, while it was predicted by Grishchuk and
Sazhin\cite{20} for their standing gravitational waves produced by
an electromagnetic generator. The determination of the amplitude
in the previous section allows to estimate the magnitude of the
shift of the test particles and opens the possibility to perform
an experimental test.
 Furthermore
it would be useful to measure also a quantity that depends neither
on the constant $A$ (and hence on the way used to fix it), nor on
the polarization tensor. To this aim we can simply compute the
direction of the shift, for example, of $m_2$ and the resulting
angle $\alpha$ is such that
$$ \frac{\delta_1}{\delta_2} = \tan\alpha = \frac{4J_1 K_1}{\lambda [2K_0^2 J_0 +
(K_0^2 + K_1^2) (J_2 - J_0)]} \tan(K_0 ct) \eqno(33)$$ and changes
quickly with time, so it cannot be useful for  a test. In the
limit $K_1 \rightarrow 0$ (the source's velocity $v \rightarrow
0$) the shift occurs in the direction of the line joining the two
particles just like the standard gravitational waves. But if we
consider the case:

d) The shift between $m_0$ and  $m_4$ at $\xi^{j} = (\ell, \ell,
0)$
\begin{figure}[ph]
\centerline{\psfig{file=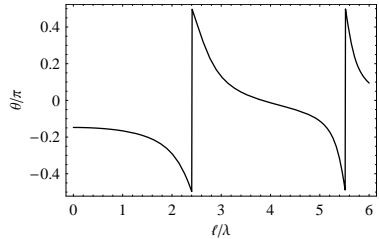,width=2.0in}} \vspace*{8pt} \caption{The "shift angle"
$\theta$ between the direction of the shift and the x - axis vs the initial position
$\ell$ of the test particle $m_4$.\protect\label{fig1}}
\end{figure}
%\begin{figure}
 % \centering
%\resizebox{\hsize}{!}{\includegraphics{Figura.eps}} \caption{The "shift angle" $\theta$
%between the direction of the shift and the x - axis vs the initial position $\ell$ of
%the test particle $m_4$.} \label{Fig. 1}
%\end{figure}

 we can compute, in the limit $K_1 \rightarrow 0$, the angle
$\theta$ in the $xy$ - plane between the direction of the shift
and the x-axis. It depends on the initial position $\ell$ of the
test particle:
$$
\frac{\delta_{2}}{\delta_{1}}=\tan\theta=-\frac{1}{2}
\left[1+\frac{J_{2}(\ell/\lambda)}{J_{0}(\ell/\lambda)}\right]
\eqno(34)$$
The dependence of the shift angle $\theta$ only on $\ell$ is a
typical feature of these waves and it is shown in Fig. 1. Of
course the angle with the line joining the two particles is
$\theta - \pi/4$ and it is different from what happens with the
standard gravitational waves. Due to the fact that $\theta$
depends neither on the time, nor on $A$, it could be used as a
constant measurable quantity in an experimental test even better
than the shift depending on $A$.
\section{Conclusions}
In some previous papers\cite{12,14,15} we studied a particular
solution (6) of the linearized Einstein equations that can play
the same role of the de Broglie wave of quantum mechanics, but we
left undetermined an arbitrary constant $A$. In this letter, we
have finally found a method to fix the constant, and hence the
amplitude of this special gravitational wave, starting from the
estimation of the corresponding energy - momentum tensor. Now we
can write down the complete solution in the form:
$$ h_{\mu\nu}= e_{\mu\nu} (K_o, K_1) \frac{4\hbar}{c^2}\sqrt{\frac{2 \pi G}{mV}} J_o\left(\frac{mc\sqrt{y^2 +
z^2}}{\hbar}\right) cos \left(\frac{Et -  px}{\hbar}\right)
\eqno(35)
$$ The result, not {\it a priori} given for granted, is very interesting because
the wave associated to the quantum particle disappears in the
limit $\hbar \rightarrow 0$ and when the source has a large mass
or volume just as it was expected. Furthermore, the knowledge of
the magnitude of the wave opens the possibility to see
experimental effects that in some cases could be more evident with
respect to the standard gravitational waves. For example, a
longitudinal shift or a direction of the shift not along the line
joining the two test particles, are peculiar effects caused by our
waves. Finally, a measurement of a particular "shift angle" that
depends only on the relative initial position of the two test
particles has been proposed.

\end{document}